# The Bubble-like Interior of the Core-Collapse Supernova Remnant Cassiopeia A

Dan Milisavljevic[1]* and Robert A. Fesen[2]

[1]Harvard-Smithsonian Center for Astrophysics, 60 Garden St., Cambridge, MA 02138.

[2]Department of Physics and Astronomy, Dartmouth College, Hanover, NH 03755.

*Corresponding author. Email: dmilisav@cfa.harvard.edu

**The death of massive stars is believed to involve aspheric explosions initiated by the collapse of an iron core. The specifics of how these catastrophic explosions proceed remain uncertain due, in part, to limited observational constraints on various processes that can introduce asymmetries deep inside the star. Here we present near-infrared observations of the young Milky Way supernova remnant Cassiopeia A, descendant of a type IIb core-collapse explosion, and a three-dimensional map of its interior, unshocked ejecta. The remnant's interior has a bubble-like morphology that smoothly connects to and helps explain the multi-ringed structures seen in the remnant's bright reverse shocked main shell of expanding debris. This internal structure may have originated from turbulent mixing processes that encouraged the development of outwardly expanding plumes of radioactive $^{56}$Ni-rich ejecta. If this is true, substantial amounts of its decay product, $^{56}$Fe, may still reside in these interior cavities.**

Computer simulations have long shown that the collapse of a high mass star's dense iron-rich core into a neutron star generates an outward moving shock wave that is unable to disrupt the star into a supernova (SN) under spherical conditions (*1*). This is because the outgoing shock wave is too weak to overcome the infalling outer layers of the star and stalls, thus requiring some post-bounce revival (*2, 3*). It is now believed that explosion asymmetries introduced by dynamical instabilities and the influences of rotation and magnetic fields must contribute to the core-collapse process (*4, 5*), but their relative contributions are uncertain. Some aspects of these explosion processes have been successfully probed by observations of extragalactic supernovae (SNe) (*6, 7*). However, even with the Hubble Space Telescope (HST), distant SNe appear as unresolved point sources, which limits our ability to unravel the three-dimensional (3D) structure of the expanding ejecta that can reveal key properties of the explosion mechanism.

An alternative approach to understanding core-collapse SN explosions is through studies of their remnants in our own Milky Way galaxy that are still young enough to be in free expansion and near enough to permit detailed studies of the expanding debris field. Such investigations can provide information on the explosion-driven mixing of the progenitor star's chemically distinct layers, the star's mass loss history before explosion, and the fate of its remnant core. With an age of around 340 years (*8*), a distance of only 11,000 light years (*9*), and its status as a known descendant of a type IIb explosion (*10*), Cassiopeia A (Cas A) is one of the best specimens for post-mortem examination. Cas A is visible today through the heating effects of a reflected or "reverse" shock generated when the original SN's high velocity blast wave ran into the surrounding interstellar medium. Cas A's metal-rich debris is arranged in large ring-like structures that together form a roughly spherical shell (commonly referred to as the main shell) approximately 6 light years in radius with an overall velocity range of −4000 to +6000 km s$^{-1}$ (*11*, *12*). These and many other properties of Cas A, such as its wide-angle (≈ 40°) opposing streams of Si- and S-rich ejecta seen in the northeast and southwest regions traveling at unusually high velocities of up to 15,000 km s$^{-1}$ (*13*, *14*), an uneven expansion of the photosphere at the time of outburst (*15*), and an overall high chemical abundance ratio of $^{44}$Ti/$^{56}$Ni with a non-uniform distribution (*16*, *17*), all point to an asymmetric explosion.

In order to investigate possible asymmetries in Cas A's interior debris, we obtained near-infrared spectra of the remnant in 2011 and 2013 using the Mayall 4m telescope at Kitt Peak National Observatory, in combination with instrumental setups particularly sensitive to the wavelength region from 900 to 1000 nm covering the [S III] 906.9- and 953.1-nm line emissions from Cas A's sulfur-rich ejecta. Closely spaced long-slit spectra were taken across the remnant's central regions (*18*). The resulting data were transformed into three-dimensional coordinates and incorporated into our existing reconstruction of the optical main shell and high-velocity outer knots (*12*).

Faint and patchy [S III] emission can be seen interior to Cas A's bright main shell with radial velocities spanning -3000 to +4000 km s$^{-1}$ in a continuous manner (Fig. 1). The emission is diffuse, in sharp contrast with the bright main shell ejecta that are compressed into small knots and filaments by the reverse shock. Most of our detections are located in the southern half of the remnant but some are found in the northern half as well (Fig. 2). Presumably, the [S III] emission arises from unshocked interior material being photoionized by ultraviolet and x-ray flux from the main shell ejecta that has been heated to temperatures up to several million degrees kelvin by the reverse shock.

The majority of the unshocked ejecta exhibit coherent structure (Fig. 3) in the form of large cavities or "bubbles" that appear to be physically connected to the main shell rings. At least two cavities are well defined. A cavity seen in the southeast with the largest range of blueshifted velocities extends outward from just below the remnant's approximate center and smoothly connects with a pair of optically bright curved filaments sometimes referred to as the "Parentheses." Opposite this, in the northwest and immediately below the remnant's largest main shell ring of redshifted material, is the largest internal cavity. These two cavities intersect along a concentration of central emission that runs from the front to the back of the remnant.

The larger northern cavity dominates the remnant's interior volume and has a radius of approximately 3 light years, whereas the southeast cavity is approximately half as large. Both cavities exhibit a few S-rich clumps or filaments indicating that neither cavity is completely empty of ejecta. Although it is difficult to accurately assess the total number of cavities, the diameters of the main shell rings are comparable to the diameters of the cavities, which suggests that they are approximately equal in number (about six). These properties, along with the fact that the main shell rings extend radially outward along gently sloped paths that follow the circumference of the cavities, support the notion that the reverse-shocked rings and unshocked cavities of ejecta share a common formation origin.

Portions of Cas A's interior unshocked ejecta that were surveyed by our near-infrared observations are also visible in previous observations taken at longer wavelengths. Infrared images of Cas A taken with the Spitzer spacecraft show [S III] 33.48 μm and [S IV] 10.51 μm emission inside the boundary of the main shell at locations coincident with regions where we detect the strongest [S III] 906.9- and 953.1-nm emissions (*19*). Follow-up Spitzer infrared spectra in the central region showed line emission from interior O, Si, and S ejecta in sheet-like structures and filaments with inferred radial velocities approaching $\pm 5000$ km s$^{-1}$ (*20*). Those spectroscopic Spitzer observations covered a smaller 50″ × 40″ area of Cas A, whereas our survey encompasses the entire remnant and reveals a far larger extent of the remnant's internal debris.

We interpret Cas A's main shell rings of ejecta to be the cross-sections of reverse shock heated cavities in the remnant's internal ejecta now made visible by our survey. A cavity filled interior is in line with prior predictions for the arrangement of expanding debris created by a post-explosion input of energy from plumes of radioactive $^{56}$Ni-rich ejecta (*21*, *22*). Such plumes can push the nuclear burning zones located around the Fe core outward, creating dense shells separating zones rich in O, S, and Si from the Ni-rich material. Compression of surrounding non-radioactive material by hot, expanding plumes of radioactive $^{56}$Ni-rich ejecta generates a "Swiss cheese" like structure that is frozen into the homologous expansion during the first few weeks after the SN explosion when the radioactive power of $^{56}$Ni is strongest.

In this scenario, the decay chain of $^{56}$Ni → $^{56}$Co → $^{56}$Fe should eventually make these bubble-like structures enriched in Fe. Doppler reconstruction of Chandra x-ray observations sensitive to Fe K emissions shows that the three most significant regions of Fe-rich ejecta are located within three of Cas A's main-shell rings (*11*, *12*). Thus, the coincidence of Fe-rich material with rings of O- and S-rich debris is consistent with the notion of $^{56}$Ni bubbles.

However, Fe-rich ejecta associated with the Ni bubble effect should be characterized by diffuse morphologies and low ionization ages, and yet the x-ray bright Fe emissions we currently see are at an advanced ionization age relative to the other elements (*23*, *24*). Furthermore, not all main shell rings have associated x-ray emitting Fe-rich ejecta, and there is no clear relationship between the locations of the internal bubbles we have detected and the spatial distribution of $^{44}$Ti recently mapped by NASA's NuSTAR (Nuclear Spectroscopic Telescope Array) (*17*).

One solution is that Fe-rich ejecta associated with some of the remnant's internal cavities remain undetected. Low-density Fe could be present in ionization states between those detectable by optical or infrared line emission and those in x-rays. The total mass of unshocked Fe that is potentially contained in the bubbles is constrained by the total nucleosynthetic yield of Fe in the original supernova explosion, which is estimated to be less than ∼ 0.2 solar mass ($M_\odot$) (*25*), and the amount of shocked Fe that is observed today, which is estimated to be 0.09 to 0.13 $M_\odot$ (*24*). Together these estimates imply that no more than an additional ∼ 0.1 $M_\odot$ of Fe could potentially be located within Cas A's reverse shock.

Whatever their true cause, Cas A's bubble-like interior and outer ring-like structures observed in the main shell suggest that large-scale mixing greatly influences the overall arrangement of ejecta in core-collapse SNe. Presently, the extent of such mixing and how it takes place are not well known. A variety of potential dynamical processes may contribute to the redistribution of chemical layers, including uneven neutrino heating, axisymmetric magnetorotational effects, and Rayleigh-Taylor and Kelvin-Helmholtz instabilities [e.g., (*26-27*)].

Compelling evidence for large-scale mixing involving considerable non-radial flow was first observed in the nearest and brightest SN seen in modern times, SN 1987A. In that case, high energy gamma-rays and x-rays with broad emission line widths from the decay of $^{56}$Ni were detected only months after the explosion, implying that Ni-rich material was near the star's surface well before 1D progenitor models had predicted assuming spherical symmetry (*28*).

Since SN 1987A, state-of-the-art 3D computer simulations of core-collapse explosions have confirmed that large-scale mixing can lead to Ni-dominated plumes overtaking the star's outer oxygen- and carbon-rich layers with velocities up to 4000 km s$^{-1}$ (*29*). However, the majority of these simulations show that the mass density should essentially be unaffected. Although mixing can affect the species distribution, the bulk of the Ni mass should remain inside the remnant with velocities below 2000 km s$^{-1}$. This is, in fact, opposite to what we currently see in Cas A, where the X-ray bright Fe has velocities around the 4000 km s$^{-1}$ limit (*11*). Thus, either the simulations are not adequately following the dynamics of mixing, or, as we suspect, more Fe remains to be detected in Cas A's interior.

An additional consideration in interpreting a SN debris field is the chemical make-up of the star at the time of outburst. The evolution of massive stars toward the ends of their life cycles is likely to be non-spherical and may produce extensive inter-shell mixing. If strong enough, these dynamical interactions lead to Rayleigh-Taylor instabilities in the progenitor structure that can contribute to the formation of Ni-rich bubbles and influence the overall progression of the explosion (*30*). Thus, asymmetries introduced by a turbulent progenitor star interior in addition to those initiated by the explosion mechanism could contribute to the bubble-like morphology observed in Cas A.

Because Cas A's opposing streams of Si- and S-rich debris have kinematic and chemical properties indicative of an origin deep within the progenitor star (*12*), we searched for evidence of any structure joining the high-velocity material with the interior ejecta mapped in our survey. However, we were not able to find any clear relationship between them. An indirect association is hinted at by the inferred projected motion of Cas A's central X-ray point source (XPS) that is thought to be the remnant neutron star. Its motion toward the southwest of the center of expansion is (i) roughly opposite to and moving away from the direction of the largest internal cavity in our reconstruction that is coincident with a sizeable concentration of reverse-shocked

Fe, and (ii) nearly perpendicular to the axis of the high-velocity jets (*31*). The XPS, conserving momentum, could have been kicked in a direction opposite the largest plume of Fe-rich material (*15*) and released an energetic protoneutron star wind that shaped the jets shortly after the core-collapse explosion (*32*).

The apparent mismatch of Ti-rich and Fe-rich ejecta regions uncovered by NuSTAR is a reminder that unresolved key issues surrounding Cas A still linger despite decades of scrutiny. Our 3D map of its interior is an important step forward as it represents a rare look at the geometry of a SN remnant's inner volume of debris unmodified by reverse shock instabilities. Because Cas A shows many striking similarities with SNe young and old (*33*, *34*), its dynamical properties described in this work are probably not unique and can be used to help interpret other SN explosions and remnants that cannot be resolved.

Our data make it clear that Cas A's dominant ejecta structure is in the form of large internal cavities whose cross-sections are the prominent rings of the reverse-shock-heated main shell. What is not clear, however, is why only half a dozen bubbles – not dozens – are present. SN explosion models can explore this issue, as well as better understand how the remnant's interior bubbles fit into a single coherent picture with its opposing high velocity jets. A crucial test of the origin of these cavities would be a confirmation of the "missing" internal Fe we predict to be located in the remnant's interior, but conclusive observations may not be possible until the next generation of infrared and x-ray space telescopes comes online.

**Acknowledgments:** This material is based upon work supported by the National Science Foundation under grant No. AST-0908237, as well as observations made with the NASA/ESA Hubble Space Telescope associated with Guest Observer program 10286 (PI: R. Fesen) and obtained from the data archive at the Space Telescope Science Institute. STScI is operated by the Association of Universities for Research in Astronomy, Inc. under NASA contract NAS 5-26555. Visual modeling of our observations was aided with the use of MeshLab (http://meshlab.sourceforge.net), a tool developed with the support of the 3D-CoForm project. We thank anonymous reviewers for providing suggestions that improved the content and presentation of the manuscript and D. Patnaude for helpful discussions.

**Supplementary Materials:**
Materials and Methods
Figure S1
Movie S1
References (*35-37*)

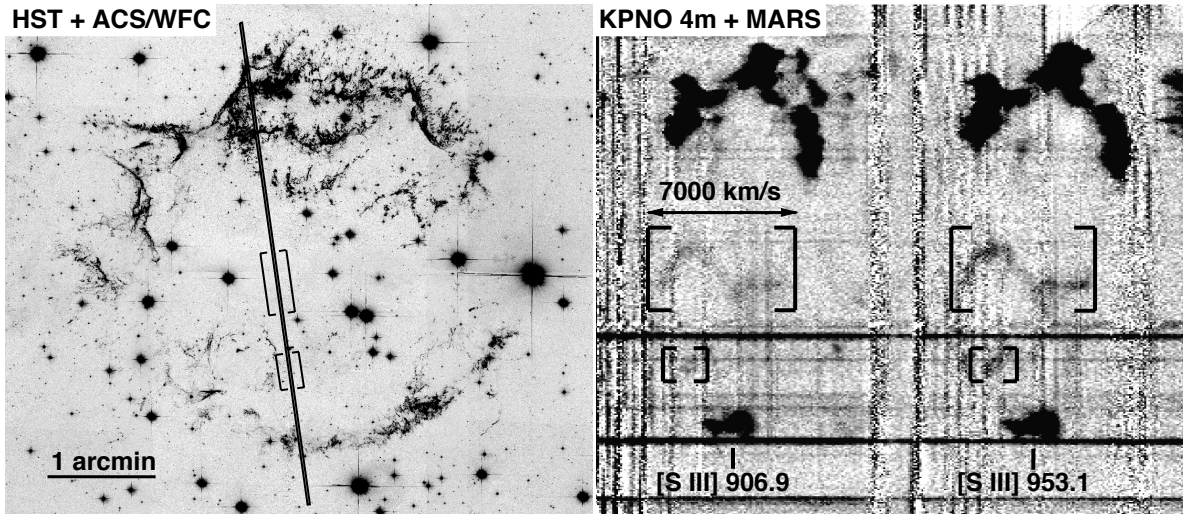

**Fig. 1. Representative near-infrared observation of the [S III] 906.9- and 953.1-nm line emission from Cas A. (Left)** A finding chart of a single long-slit position, which was rotated to optimize coverage of particular interior emission regions. Background image is a mosaic created from 2004 HST observations sensitive to oxygen and sulfur emissions (*31*). ACS/WFC, Advanced Camera for Surveys/Wide Field Channel. **(Right)** The corresponding 2D spectrum. Square brackets highlight regions where interior unshocked ejecta have been detected. See fig. S1 for a map of all slit positions used in the survey. KPNO, Kitt Peak National Observatory; MARS, Multi-Aperture Red Spectrograph.

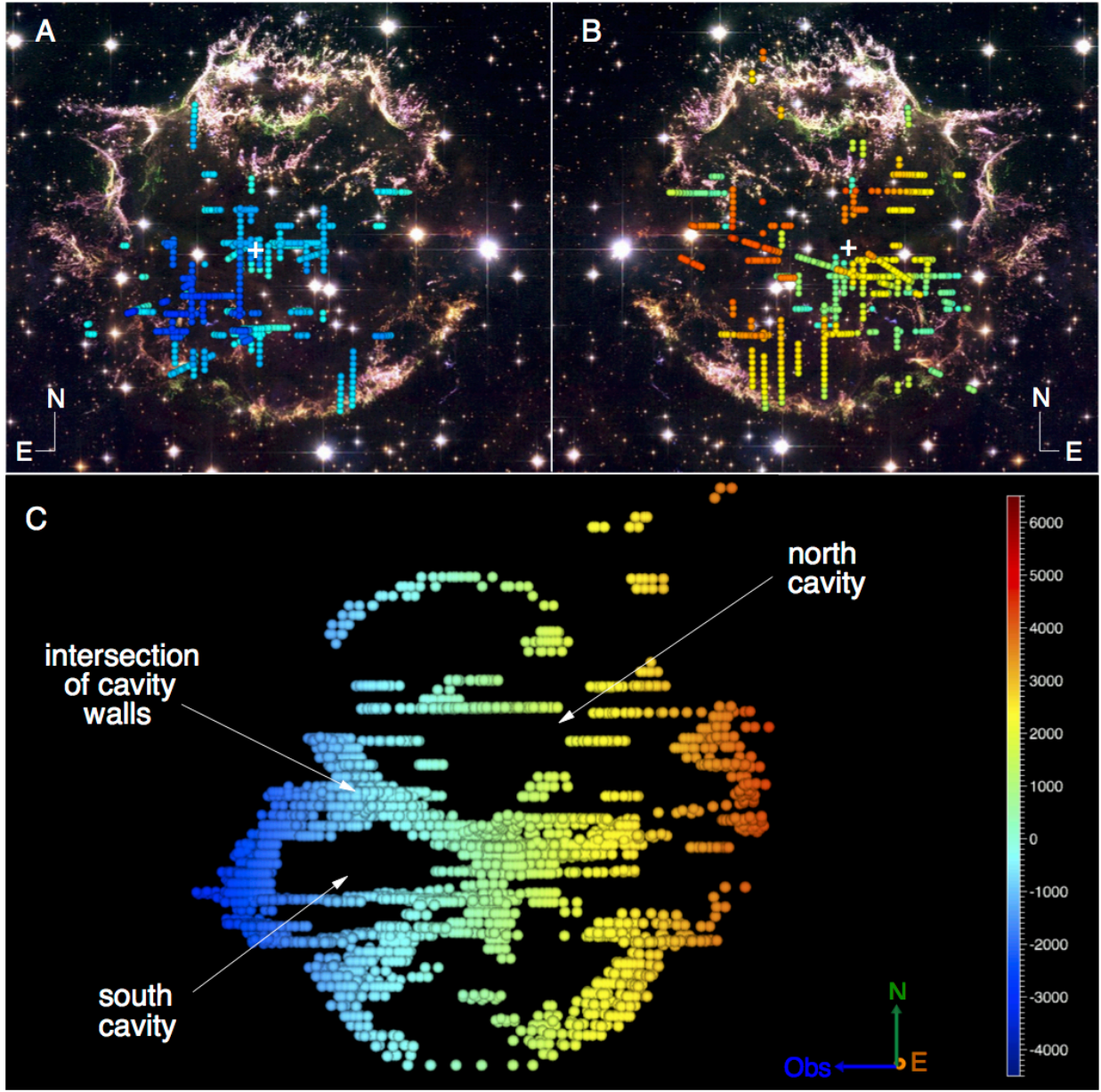

**Fig. 2. Map of [S III] 906.9- and 953.1-nm line emission detected by our survey.** The blue-to-red color gradient represents the range of measured Doppler velocities. Spheres mark individual measurements. (**A**) All blueshifted emission ( < 0 km s$^{-1}$) and (**B**) all redshifted emission (> 0 km s$^{-1}$). The background is a composite HST image sensitive to the remnant's oxygen and sulfur emissions retrieved from www.spacetelescope.org. The center of expansion is shown as a white cross. (**C**) A perspective rotated 90° toward the west along the north-south axis. The north and south interior cavities are highlighted, as well as the wall of ejecta where the two cavities intersect.

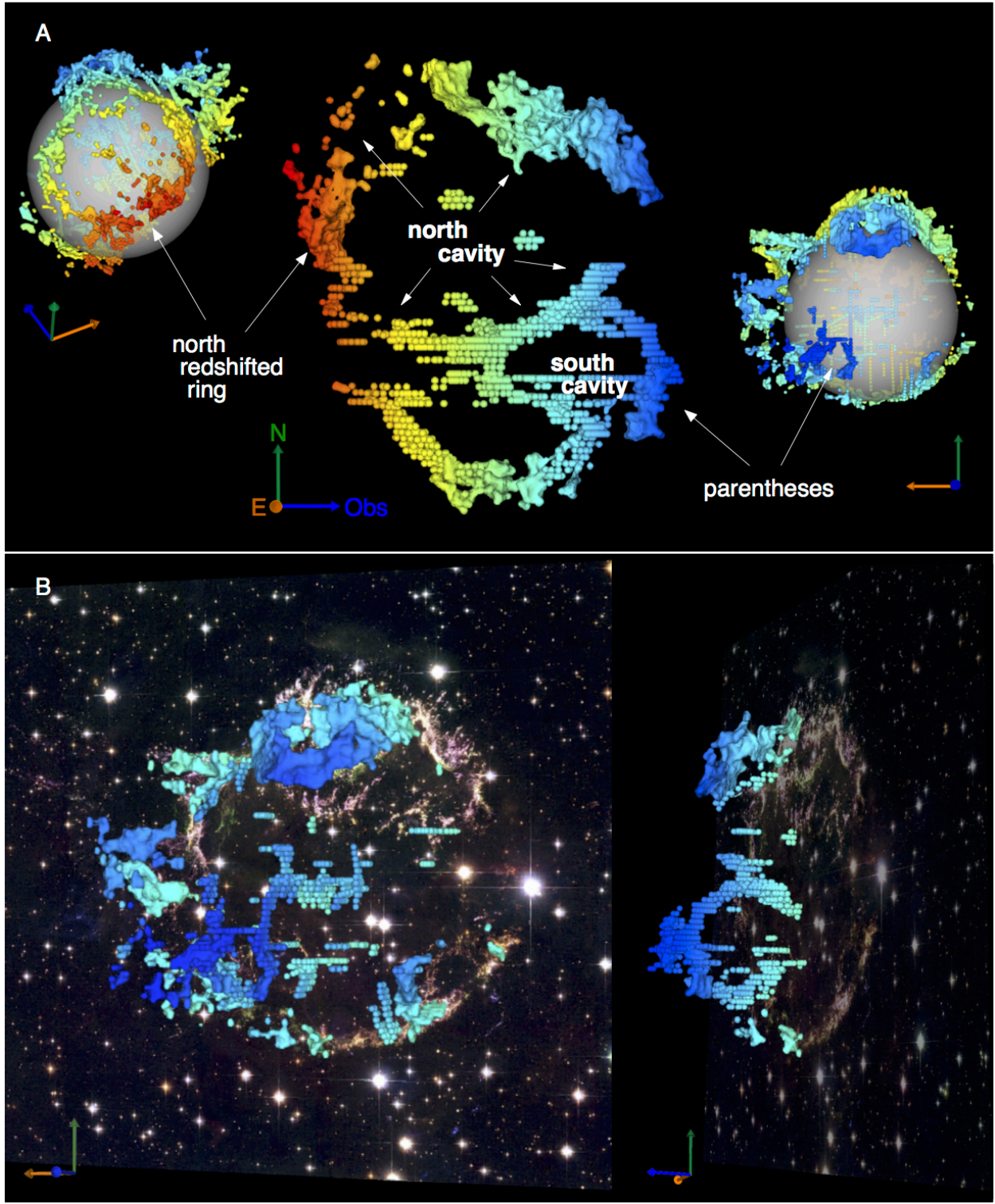

**Fig. 3. Doppler reconstruction of Cas A made from the [S III] 906.9, 953.1 nm emission map presented here and previous optical observations of main shell ejecta (*12*).** The blue-to-

red color gradient corresponds to Doppler velocities that range from -4000 to 6000 km s$^{-1}$. [S III] measurements are individual spheres, and previous optical data are smoothed with a surface reconstruction. (**A**) A side perspective of a portion of the remnant spanning all material located between 15″ east of the center of expansion to 50″ west of the center of expansion to emphasize the two conspicuous interior cavities and their connections to main shell ejecta. The translucent sphere centered on the center of expansion is a visual aid to differentiate between front and back material. (**B**) Two angled perspectives highlighting the south cavity. The first perspective angled 20 degrees away from the observers line of sight shows all data, and the second perspective angled 70 degrees away from the observers line of sight shows the same portion of the remnant as displayed in Panel A. The background image representing the plane of the sky as seen from Earth is the same shown in Fig. 2. An animation of the entire reconstruction is provided in movie S1.

# Supplementary Materials

## S1. Spectroscopy

Long-slit spectroscopy of the Cassiopeia A supernova remnant's interior was obtained 2011 September 19–22 using the Mayall 4m telescope located at Kitt Peak National Observatory in Arizona. The Multi-Aperture Red Spectrometer (MARS) instrument was used in combination with a red-sensitive LBNL CCD detector and VPH 8050-450 grism and GG495 order blocking filter. Spectra were obtained using a 1.2″ slit and covered the wavelength region of 5500–10800 Å with full-width-half-maximum resolution of ≈8 Å.

Supporting observations were also obtained in 2013 September 30–October 4 with the Mayall 4m telescope but this time with the RC Spectrograph instrument in combination with the LB1A CCD detector and BL 181 grating and OG570 order blocking filter. A larger 2″ slit was used and covered the wavelength region of 6300-9800 Å with full-width-half-maximum resolution of ≈8 Å. The 2013 observations did not achieve the depth of the earlier MARS observations but did confirm the prior observations of the 2011 run and complemented the survey.

Figure S1 shows a finding chart of all 40 positions that were obtained. The positions followed a loose grid pattern that was weighted toward [S III] emission detected in previous infrared images. Exploratory spectra that were at angled positions were also included in the final analysis. At each position, 2-3 exposures between 1200-1800 seconds in length were obtained. Observations were made below an airmass of 1.4 to minimize effects of atmospheric refraction.

Comparison He–Ne–Ar lamp images were taken at regular intervals between positions for wavelength calibration. Stars with well-measured coordinates encountered during the progression of long-slit positions provided fiducial reference points from which to ensure consistent and accurate coordinates. In some cases, bright stars were intentionally observed to provide stellar continua by which accurate positional off-sets could be determined. Regions of overlap between various long-slit positions were compared to check for detections of emission. These regions also provided additional checks on the coordinates.

Data were reduced using standard procedures in IRAF/PyRAF. The 2D images of each position were trimmed, bias-subtracted, flattened, and co-added to remove cosmic rays. Images were then wavelength calibrated in the dispersion axis using the comparison lamp images and straightened in the spatial axis using tracings of stellar continua. The sky background of the images was subtracted with the IRAF task *background* using a fifth-order Chebyshev function fit along the spatial direction that was sampled from the median of 50 pixel bins.

## S2. Data Analysis and Doppler Reconstruction

The final images were inspected by automated and manual procedures for emission associated with the [S III] 906.9, 953.1 nm lines. Signal was recorded as a positive identification if emission could be detected in both of the doublet lines with a signal-to-noise ratio of at least 2.5 above the background. We note that some of the brighter S-rich material was also marginally detected in [S II] 671.6, 673.1 nm emission lines, but because near-infrared emission is less sensitive to line of sight extinction that amounts to 5-8 magnitudes at visible wavelengths (35), the emission is more easily observed in the near-infrared [S III] 906.9, 953.1 nm lines. These detections, nonetheless, provided additional confidence that the emission was real and the Doppler shifts measured accurate.

The spatial locations of verified detections were associated with coordinates in right ascension (RA) and declination (DEC), and their wavelength positions converted to Doppler velocities. Recorded positional coordinates are believed to be accurate to within 1.2″ (of order the slit width), and Doppler velocities accurate to within ±100 km/s.

Over 50 years of optical monitoring of Cas A have demonstrated that its S- and O-rich knots have undergone velocity changes of less than 5% over 300 yr (*8, 36, 37*). We assume that the inner material follow similar ballistic trajectories from the well-constrained center of expansion (COE) that is located at RA(2000.0) = $23^h23^m27^s.77$ and DEC(2000.0) = +58°48′49″.4 (*8*). This enables us to relate radial velocities into line-of-sight distances from the COE, and transform RA and DEC coordinates into a velocity representation by dividing distance from the COE with the conversion factor of 0.022″ per km s$^{-1}$ [see (*12*) for details].

For the animation Movie S1 a surface reconstruction has been performed on the resulting point cloud using a ball-pivoting algorithm to interpolate and smooth the 3224 individual measurements of the [S III] emissions detected in our survey. The surface layer was created 170 km s$^{-1}$ above the individual measurements in order to blend the non-uniform sampling of the slit positions. The surface reconstruction mimics the diffuse nature of the emission and guides the eye in locating coherent structure. We also show the individual measurements from which the surface reconstruction was made to explicitly define what was measured and what has been enhanced.

## S3. Cassiopeia A Web App

A web application powered by Javascript/WebGL has been developed to allow users to interact with a simplified version of the 3D reconstruction of the supernova remnant Cassiopeia

A presented in this work. The app is compatible with most modern browsers and can be found at this internet address:

https://www.cfa.harvard.edu/~dmilisav/casa-webapp/

## References (35-37)

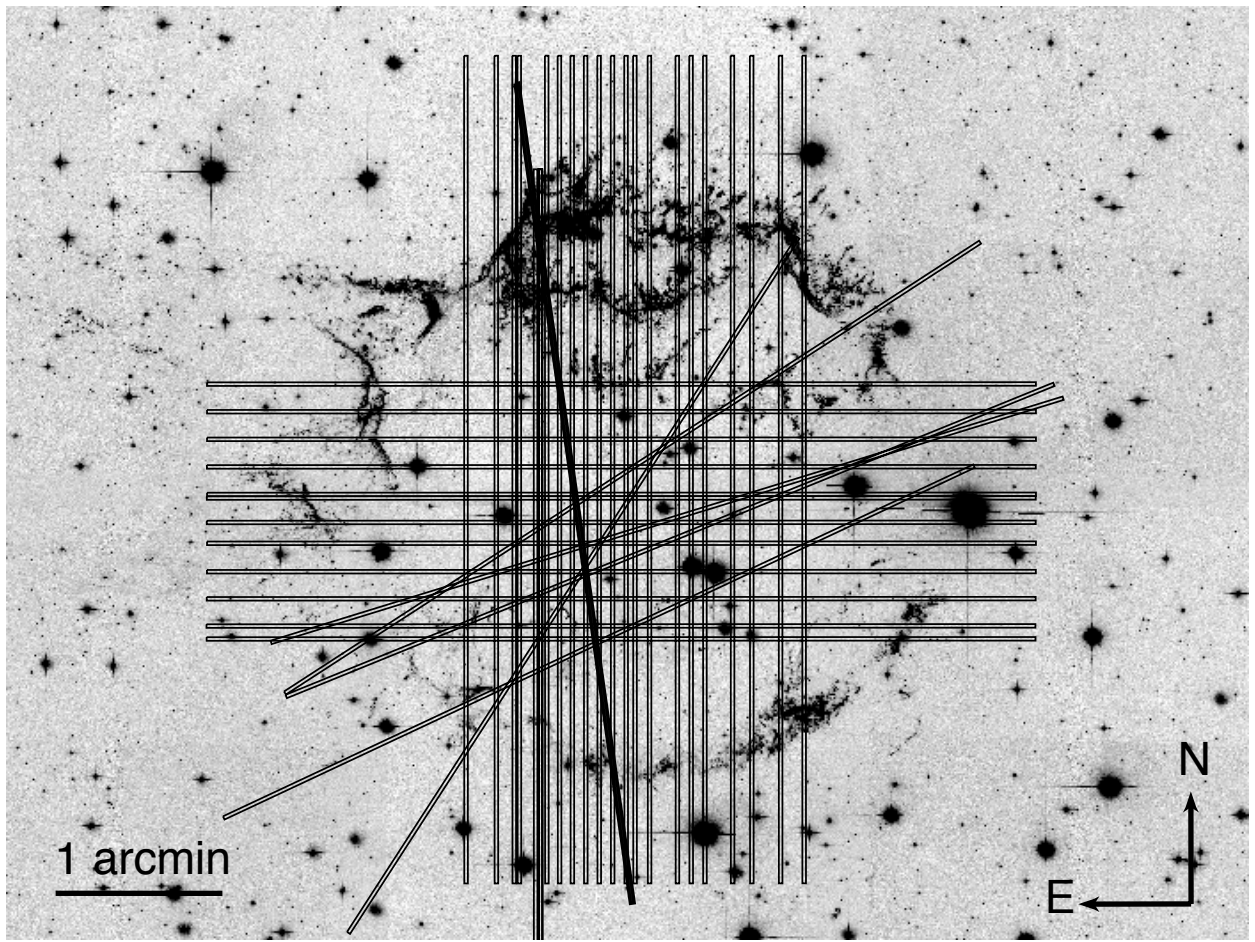

**Fig. S1.**

Finding chart of slit positions obtained in our survey. Background image is a mosaic created from 2004 HST/ACS observations sensitive to oxygen and sulfur emissions (*31*).

**Movie S1**

3D Doppler reconstruction of Cassiopeia A. First shown are individual measurements of [S III] 906.9, 953.1 nm emission represented as spheres. The main shell is then shown using data from (*12*). Relevant features of the main shell that connect with the interior cavities are highlighted. A translucent sphere centered on the center of expansion is a visual aid to differentiate between front and back material. The individual measurements are then replaced with the surface reconstruction. The south cavity that connects with the "Parentheses" and north cavity that connects with the north, redshifted ring are highlighted, as well as the region of central emission where the two cavities intersect. Finally, the entire reconstruction is shown with the backdrop of an HST image (retrieved from www.spacetelescope.org) representing the plane of the sky. Color gradient blue-to-red corresponds to Doppler velocities that range from -4000 to 6000 km s$^{-1}$, and the central sphere in white is the center of expansion.